\documentclass[english]{article}

\usepackage{geometry} 
\usepackage{graphicx}
\usepackage{amssymb}
\usepackage{epstopdf}

\usepackage{babel}
\usepackage{inputenc}
\usepackage{epsfig}

\begin{document}

\begin{flushright} 
\today \\

\end{flushright}

\vspace{0.8cm}

\begin{center}
{\Large {\bf Rising Total Hadron-Hadron Cross Sections }}
\end{center}


\begin{center}
\normalsize{
Giorgio Giacomelli
\par~\par

Dipartimento di Fisica dell'Universit\'a di Bologna and 
INFN Sezione di Bologna, I-40127 Bologna, Italy\\ 
giacomelli@bo.infn.it} 

\par~\par

\vspace{0.5cm}
{\bf In honour of George T. Zatsepin}

\end{center}

\vspace{0.5cm}

{\bf Abstract.} {\normalsize A historical summary is made on the measurements concerning the 
rising total hadron-hadron cross sections at high energies. The first part of this paper concerns the total cross section measurements performed at the Brookhaven, Serpukhov 
and Fermilab fixed target accelerators; then the measurements at the CERN Intersecting Storage Rings (ISR), and at the CERN and at the Tevatron Fermilab $\bar pp$ colliders; finally the cosmic ray measurements at even higher energies. A short discussion on Conclusions and Perspectives follows.

}

\section{Introduction} Hadron-hadron total cross sections were accurately 
measured at most new hadron accelerators which opened up new energy regions. 
Most of the systematic total cross section measurements of the 6 long-lived 
charged hadrons ($\pi^{\pm}, K^{\pm}, p^{\pm}$) on hydrogen and deuterium 
targets at fixed target accelerators were performed using the transmission 
method, pioneered in the 1960's at Brookhaven National Laboratory (BNL); 
the method is capable of high precisions, typically point to point precisions 
of $\sim$0.2\% and a systematic scale uncertainty of $<$1.0\%.\par

Fig. 1, from the Data Particle Group, shows the behaviour with energy of the 
total cross sections  of $\pi^{\pm}p$, $\pi^{\pm}d$. At low energies, in the so called 
{\it resonance region}, one observes a number of peaks and structures which 
decrease in size as the energy increases. Above 5 GeV/c lab momentum, in the 
{\it continuum region}, there are no more structures: the cross sections 
decrease smoothly, reach a minimum and then slowly rise with increasing energy 
(the {\it asymptotic region}). In the low energy region the cross sections 
depend strongly on the type of colliding hadrons and on the total isotopic 
spin, while at high energies these dependences tend to disappear as the energy 
increases.\par

The BNL experiments  in the 1960's concerned the resonance region and the beginning of the 
continuum region. The experiments performed later at the then new Serpukhov 
accelerator by the CERN-Serpukhov collaboration in the early 1970's discovered the flattening of the $K^-$p  
total cross sections (1970) and the rising $K^+$p 
total cross sections (1971). Then followed the experiments at the CERN ISR $pp$ 
collider, where it was found that the pp total cross section was also rising (1973). 
Later  in the 1970's systematic measurements were made at the new Fermilab Tevatron fixed 
target accelerator using the BNL-Serpukhov method: it was found that also the 
$\pi^{\pm}$p and $K^-$p total cross sections were rising with increasing 
energy. One had to wait for the CERN and Tevatron $\bar p p$ colliders to 
prove that also the $\bar p p$ total cross section was rising.\par

The highest energies were and are still available only in cosmic rays: cosmic ray (CR) measurements indicate in 1972 that the pp total cross section was rising as the energy increased.

\begin{figure}[!h]
 \centering
 {\centering\resizebox*{!}{13cm}{\includegraphics{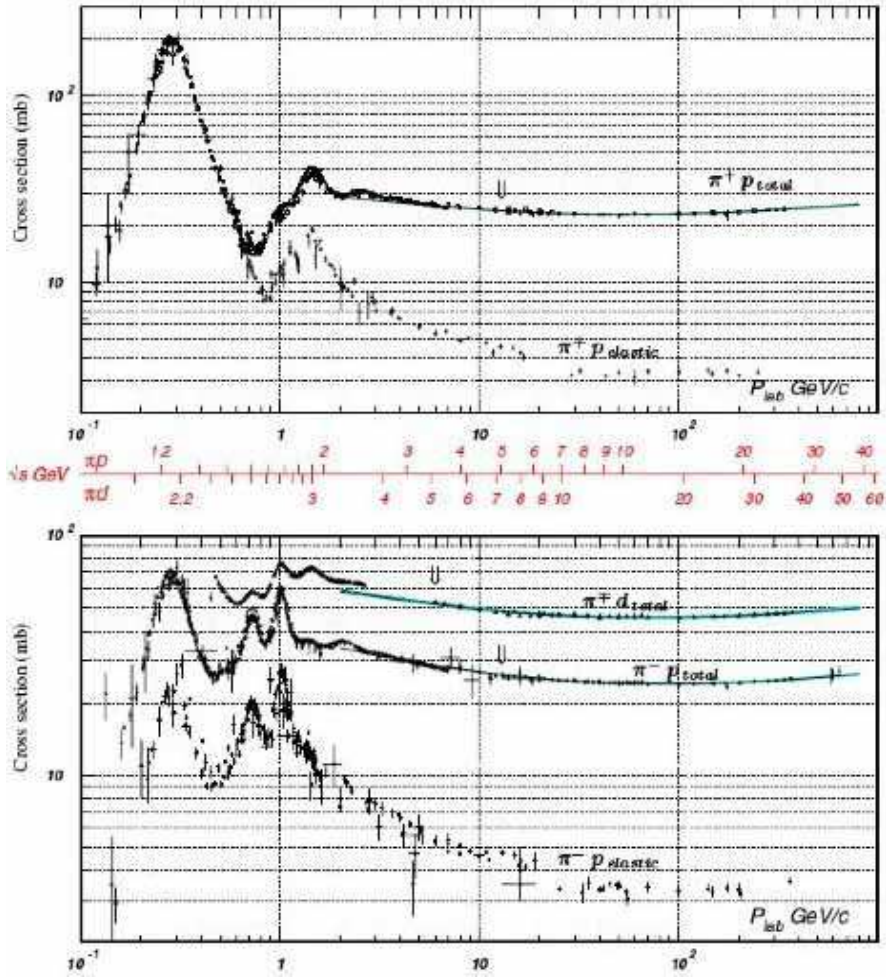}}\par}
\begin{quote}
 \caption{\small Compilations of the total and of the integrated elastic cross 
section data versus lab momentum for $\pi^{\pm}$p and $\pi^{\pm}$d scattering 
[Particle Data Group, 2006]. } 
\label{fig:plots}
\end{quote}
 \end{figure}

\section{Total cross sections at the Brookhaven 33 GeV AGS} 
At BNL a series of measurements were made with different beams covering the resonance 
region and the beginning of the continuum region, 0.5$< p_{lab}<$22 GeV/c  \cite{diddens, cool, abrams}. \par

A precise total cross section measurement in the resonance region was then a method 
to detect new resonances and this was the main aim of the Brookhaven measurements. 
Low mass resonances are easy to detect because they produce large effects. Higher 
mass resonances show up as broad and non prominent structures, often 
overlapping with one another, so that one needs to measure the total cross sections 
with high precision at many closely spaced points. Errors in the absolute values 
can be tolerated if they are essentially energy independent. \par

The $\pi^+$p, $K^+$p, and pp are pure isospin states. In the other cases one has a 
mixture of two isospin states. The determination of the pure isospin cross sections 
requires the measurement of two cross sections, which involves changing either the 
incident or the target particle. For pions it is easy to measure both $\pi^+$p 
and $\pi^-$p cross sections, and hence to derive the total cross sections 
$\sigma_{1/2}$ and $\sigma_{3/2}$ for pure isospin states. For the other cases the 
simplest solution is to measure the cross sections off protons and off neutrons. 
The best neutron target is a bound neutron-proton state (the deuteron): problems of 
nuclear physics in the deuteron limit the analysis of the data and an unfolding 
procedure must be performed to extract the pure isospin cross sections.\par

Total cross section measurements do not provide enough information to establish 
conclusively that a peak in a definite isospin state is a resonant state, i.e. a 
state with definite quantum numbers. In fact, a structure could also come from a 
threshold effect, such as the opening up of a new important channel, 
or other kinematical effects. \par

The method employed was that of a standard 
transmission ``good geometry'' experiment. 
The used low energy beams were partially separated secondary beams. 
After momentum and mass separation, the beam was defined by a sysytem of scintillation 
counters and by a Cherenkov counter, which electronically distinguishes 
between wanted and unwanted particles. The beam alternatively passed 
through a hydrogen, deuterium, or dummy target and converged to a focus 
at the location of the transmission counters, each of which subtended a different 
solid angle from the center of the target. This allowed to 
evaluate the partial cross sections  $\sigma_i$ measured by each individual 
transmission counter and to extrapolate these cross sections to zero solid angle 
to obtain the total cross section.\par

In the $K^+$N, I=0 state there is a structure at the center of mass (c.m.) energy of 
about 1910 MeV. Many measurements were made on this system, without 
reaching a final conclusion, though a possible I=0 resonant state seemed to be 
indicated for this ``exotic system'' \cite{gg}. The $K^+$p system received considerable attention few years ago, with the possible observation of a narrow ``pentaquark state''. This 
possibility seems now to be disfavoured \cite{oh}.

\section{Total Cross Sections at the IHEP 70 Gev protonsynchrotron} 
The program of the first CERN-IHEP (Serpukhov) Collaboration concerned the measurement 
of the energy dependence, first in 1969 of the $\pi^-$p, $K^-$p and $\bar p p$ total cross 
sections  and later, in 1971, of the $K^+$p, $\pi^+$p and pp total cross sections in the lab 
momentum range 15-60 GeV/c \cite{allaby, denisov}. The experimental method was similar to that used in Brookhaven, that is a standard 
transmission method in good geometry, using more refined Cherenkov counters (see the similar layout used at Fermilab, Fig. 
\ref{fig:apparatus})

The 1969 results from the first set of measurements with negative particles ($\pi^-$, $K^-$, $\bar p$) indicated that the decrease of the three measured cross 
sections almost stopped, leading to essentially energy independent total 
cross sections, Figs. 1, 4. The results from the second (1971) set of measurements using 
positive particles ($p^+$, $K^+$, p) in the same momentum range lead to similar conclusions for $\pi^+$p and $pp$, Figs. 1, 4, and to the surprising 
discovery of rising $K^+$p,  $K^+$d total cross sections, Figs. 1, 2, 4. This came as a surprise to 
most physicists\footnote{Personal recollection. At the beginning of 1970 a 
group of CERN physicists involved in the first CERN-Serpukhov experiment, 
before leaving for a new run period at Serpukhov, had a coffee discussion in 
the CERN canteen. They were joined by several friends. The discussion 
concerned the expected asymptotic energy behaviour of the total hadron-hadron 
cross sections: most experimentalists favoured constant, energy independent 
cross sections, while most theoreticians favoured decreasing cross sections 
going towards zero. In the middle of the discussion arrived Giuseppe Cocconi, 
the CERN ``father'' of these types of measurements: he listened for a while, 
then he ``exploded'': ``It is all nonsense: I bet a coffee that the cross 
sections will rise!'' This proposal sounded a bit crazy, so I and others 
accepted the bet... and a couple of years later, at the beginning of 1972, 
Cocconi wanted the free coffee! }, even if some theoreticians had predicted a possible rise \cite{cheng}.  Fig. 2 shows the rising $K^+$p total cross section at increasing energies measured at Serpukhov; the same features were observed in the $K^+$d and $K^+$n cross sections. The $\pi ^+$p and $pp$ data were instead found to be almost energy independent, suggesting that they had a minimum at these energies, Fig. 2, 4. Moreover $\sigma_{tot}(pp) \simeq \sigma_{tot} (pn)$ in agreement with isospin independence. The comparison of the total cross sections for particles and antiparticles on protons indicated that their differences were decreasing with increasing energies. This was particulary evident when plotting the total cross section differences:

\begin{equation}
\Delta \sigma = \sigma_{tot} (\bar x p) - \sigma_{tot}(xp) = A~ p_{lab} ^{-n}
\end{equation}

\noindent The behaviour is consistent with the Pomeranchuck theorem according to which $\Delta \sigma \rightarrow 0$ as $p \rightarrow \infty$ \cite{okun}. \par

The study of charged hadron production vs lab momentum was also an important point, as indicated in ref. \cite{baker, gg2}. Also the measurements of the absorption cross sections in various nuclei was a relevant point \cite{binon}. \par

In the subsequent years there were cosmic ray experiments which indicated a possible increase of $\sigma_{tot}(pp)$ at the highest Cosmic Ray (CR) energies \cite{yodh}.

\begin{figure}[!h]
\centering
{\centering\resizebox*{!}{4.5cm}{\includegraphics{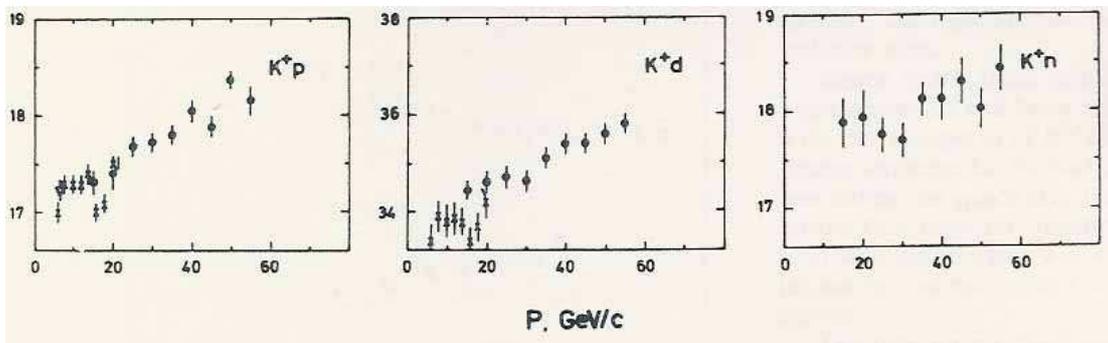}}\par}
\begin{quote}
\caption{\small The rising K$^+$p, K$^+$d and K$^+$n total cross sections measured at Serpukhov by the CERN Serpukhov Collaboration 
\cite{denisov} } 
\label{fig:fig2}
\end{quote}
\end{figure}

\section{Total Cross Sections at the CERN-ISR} 
The CERN Intersecting Storage Rings (ISR) consisted of two concentric and slightly distorted rings for protons, each 300 m in diameter. The two rings intersected horizontally eight times, with a crossing angle of 14.8$^{\circ}$. The ISR operated at c.m. energies $\sqrt{s}$  = 23.4, 30.4, 44.4, 52.6 and 62.3 GeV \cite{amaldi, gg3, gg-jacob}. A key parameter of the ISR was the luminosity L, which determined the total number of interactions per unit time, R, in each intersection

\begin{equation}
R = L \sigma_{tot} (pp)
\end{equation}

\noindent where $\sigma_{tot} (pp)$ is the cross section at each energy. The luminosity was measured by the Van der Meer method of displacing the two beams vertically from one another, recording the rate R in a monitor \cite{amaldi, gg3, gg-jacob}. \par

The direct method for the measurement of the $pp$ total cross 
section at the ISR was based on the application of the luminosity 
formula, Eq. 2, performing separate measurements of R 
and L. The total number of interactions was measured with large scintillation 
counters; extrapolations had to be made to take into account the missing 
number of interactions, mainly at small angles. L was 
measured as stated above.  \par

Indirect methods for the measurement of $\sigma_{tot} (pp)$ were connected with the 
use of the optical thorem
\begin{equation}
lm ~F(s,0)= s ~\sigma_{tot}(s)
\end{equation}

\noindent Squaring expression (3) leads to a relation between the square of the total 
cross section and the elastic differential cross section at t=0

\begin{equation}
\sigma^2 _{tot}= 16 \pi (\hbar c)^2 (dN_{el}/dt) |_{t=0} /L (1+\rho^2)
\end{equation}

\noindent where $(\rho = Re F/lm F)|_{t=0}$ is the ratio at t=0. The differential elastic cross section is written as

\begin{equation}
dN_{el}/dt = |dN_{el}/dt |_{t=0}~ exp(-B|t|)
\end{equation}

Another expression for the total cross section is

\begin{equation}
\sigma_{tot} = (N_{el}+N_{inel})/L
\end{equation}

\noindent $N_{el}$ and $N_{inel}$ were measured simultaneously. Three indirect methods where used at the ISR.\par

The first method used the measurement of the elastic cross section at small 
angles, extrapolating it to t=0 by means of Eq. 3, then using Eq. 4 
with a measurement of the luminosity L, and an estimate for 
$\rho = Re F/Im F$, assumed to be t-independent.\par
The second method was based on the measurement of the elastic scattering 
cross sections in the Coulomb-Nuclear interference region, for 
$0.001<|t|<0.01$ (GeV/c)$^2$. In this region the expression for the cross 
section depends on the high energy parameters $\rho$, B, $\sigma_{tot}$. 
Several types of fits were performed, for example leaving both $\rho$ and 
$\sigma_{tot}$ as free parameters. In this case one has an absolute 
normalization to the Coulomb scattering formula, which is well calculable.\par

The third method was based on the simultaneous measurements of the total 
collision rate and of elastic scattering in the nuclear region, then using 
Eq. 3: the measurement of $\sigma_{tot}$ does not depend on 
the luminosity, thus removing one of the uncertainties. \par

All measurements indicated the $\sigma_{tot} (pp)$ was rising with increasing energies, see Fig. 4.

\section{Total Cross Sections at the Fermilab fix target accelerator}
Fig. 3 shows the layout of the total cross section measurements at the Fermilab fix target accelerator (separated function synchrotron operating at 300 and at 400 GeV). 
The differences compared to previous measurements were due to 
the higher energies of the Fermilab beams, thus to the need of more selective 
differential Cherenkov counters. Incident particles were 
defined by scintillation counters and identified by two differential gas 
Cherenkov counters, allowing cross sections of two different particles to 
be measured simultaneously. In addition, a threshold gas Cherenkov counter 
could be used in anticoincidence. Sufficient $\pi^+ - K^+$ separation was achieved up to 200 GeV/c, 
and at higher momenta using corrected optics \cite{carroll2}. Contamination of unwanted 
particles in the selected beam particles was $<$0.1\%. 
In the pion and kaon beams there were small admixtures of muons and electrons 
(at the level of 0.1\% and 1\%, respectively). Electrons were 
identified by their characteristic signal in a 22-radiation length lead-glass 
Cherenkov counter placed downstream of the transmission counters. 
Muons were identified by their ability  to pass through 5 m of steel 
placed downstream of the transmission counters. Other differences concerned 
the order in the transmission counters (first the large transmission counters at 
Brookhaven, and the reverse at Fermilab). The transmission counters could be moved on 
rails so  as to subtend at each energy the same t-range. The data were taken first in the range 50 to 200 GeV/c secondary beam momentum, and 
later in the ranges 23-280 and 200-370 GeV/c.  \par

\begin{figure}[!h]
 \centering
 {\centering\resizebox*{!}{6 cm}{\includegraphics{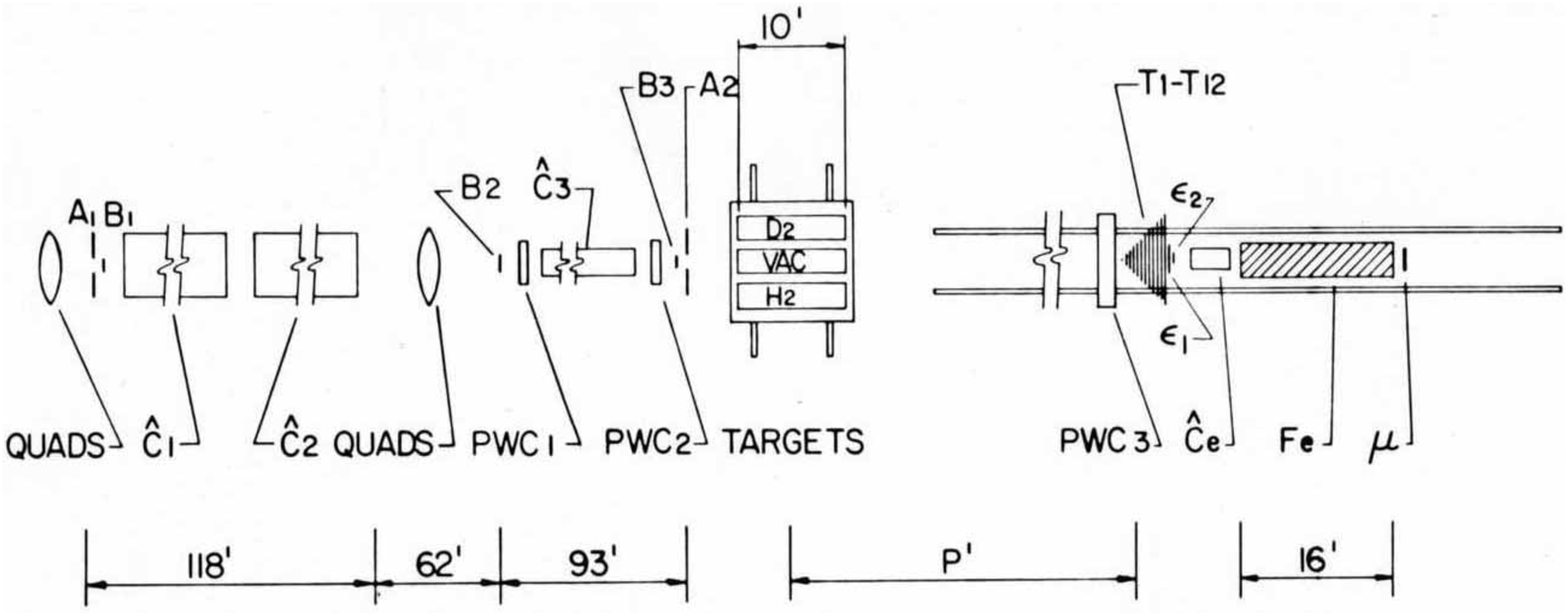}}\par}
\begin{quote}
 \caption{\small Layout of the apparatus for the measurement of the charged hadron total cross sections at Fermilab. 
$C_1$, $C_2$ $, C_3$ are gas differential (threshold) Cherenkov 
counters, PWC1-PWC3 are proportional wire chambers $B_1$-$B_3$ and 
$A_1$-$A_2$ are scintillation counters. $H_2$, $D_2$, VAC are the liquid 
hydrogen liquid deuterium and dummy targets, 
$T_1$-$T_{12}$ are transmission counters, $\epsilon_1$-$\epsilon_2$ 
are small scintillation counters used for efficiency measurements, 
$C_e$ is a lead glass Cherenkov counter; the iron absorber and the muon ($\mu$) scintillation counter were used to estimate the muon 
contamination. The beam was counted as $B=B_{123}~ \bar A_{12}~ C_i$, where 
$C_i$ was a combination of the 3 Cherenkov counters. } 
\label{fig:apparatus}
\end{quote}
 \end{figure}

A compilation of all measured data is given in Fig. 4a. 
These measurements reveal that the total cross sections and thus the effective sizes of 
both the proton and neutron increase for five of the six probes when their lab 
energy increases. For the sixth, the antiproton, 
the rapid decrease previously observed below 50 GeV/c had slowed down and 
the apparent size becomes essentially constant above 120 GeV/c.\par

The similarities of the behaviour of the cross 
sections with the six probing particle beams indicate that a new simplicity 
of nature was revealing itself at high energies. All of the particle-proton and antiparticle-proton cross section pairs 
uniformly approach each other, see Figs. 4, 5.  For each probe particle, the neutron cross section is nearly equal to 
the proton cross section. The differences between particle and antiparticle 
pairs seems to be disappearing at very high energies. \par

As already stated, the study of total cross sections requires first a study of the beam qualities 
and of their fluxes; this provides interesting information on the 
production cross sections of the six long lived charged hadrons, see Fig. 6 \cite{baker}. Besides the liquid hydrogen, deuterium and dummy targets, one had 
always available a number of targets of different materials 
(Li, C, Al, Cu, Sn and Pb). Thus one had the possibility of measuring the 
absorption cross sections in nuclei  \cite{binon}.\par

The multiplicity of charged particles produced in inelastic processes was measured in several experiments and was found to be increasing with energy \cite{antinucci}.

\begin{figure}[!h]
 \centering
 {\centering\resizebox*{!}{7.8cm}{\includegraphics{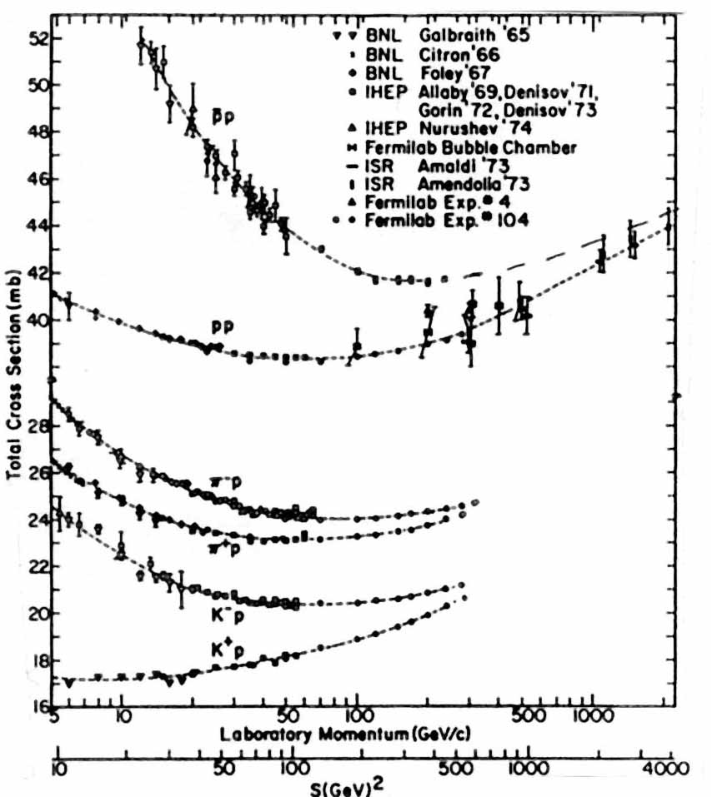}}}
 {\centering\resizebox*{!}{7.8cm}{\includegraphics{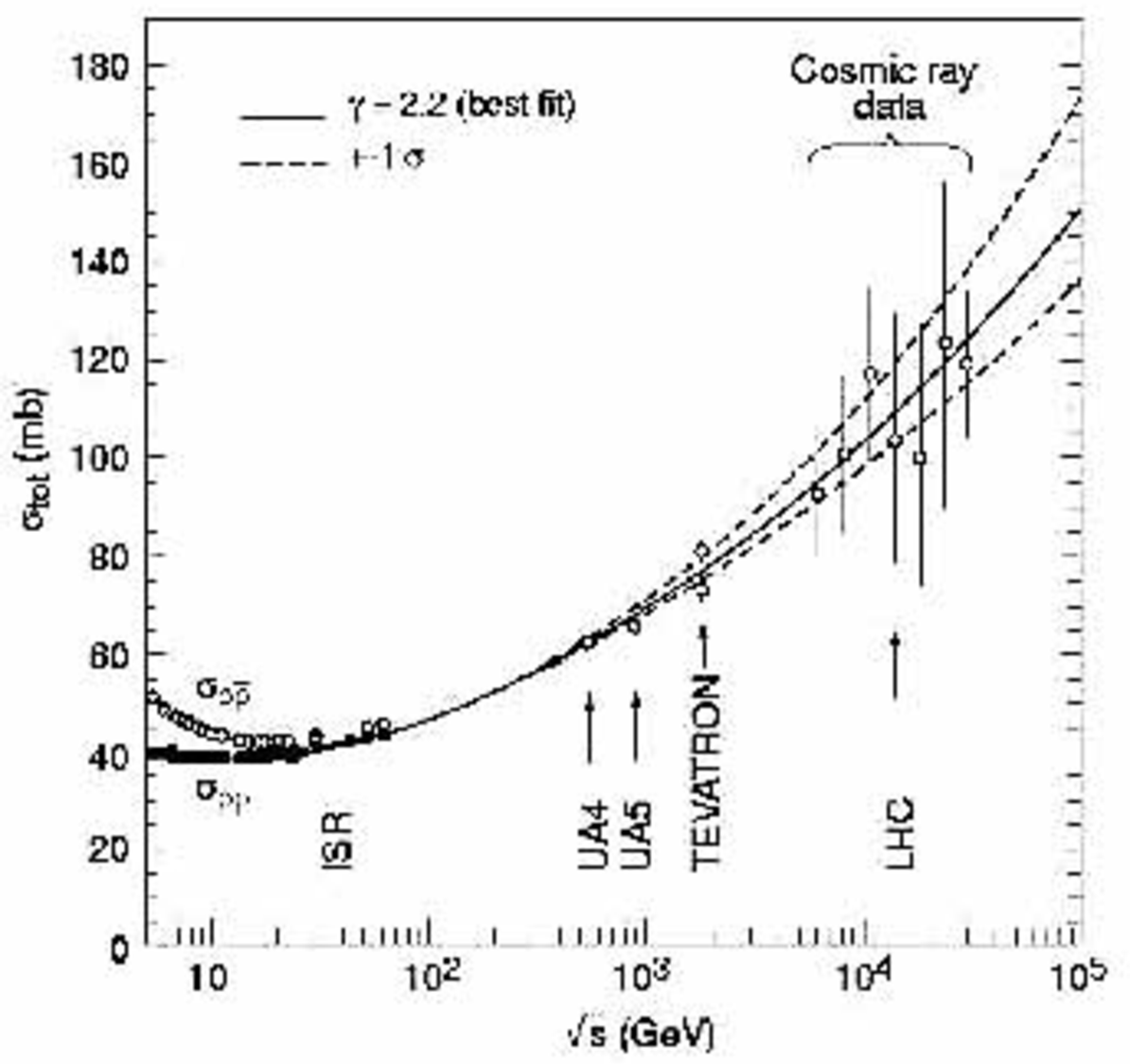}}\par}
\begin{quote}
 \caption{\small (a) Compilation of $\bar p p$, $pp$, $\pi^- p$, $\pi^+ p$, 
$K^- p$, $K^+ p$ total cross sections plotted versus c.m. energy. 
(b) The $\bar p p$ and the $p p$ total cross sections, including cosmic ray 
measurements. The solid line is a fit of the $\sigma_{tot}$ and $\rho$ 
data with dispersion relations; the region of uncertainty is delimited by dashed lines. } 
\label{fig:compilation}
\end{quote}
 \end{figure}

\begin{figure}[!h]
 \centering
 {\centering\resizebox*{!}{8cm}{\includegraphics{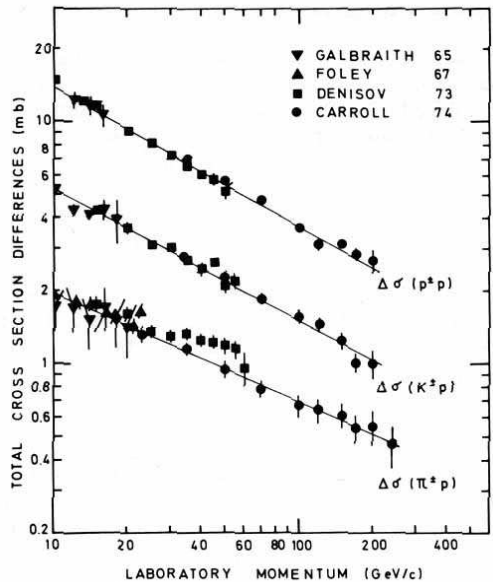}}\par}
\begin{quote}
 \caption{\small  The differences of total cross sections for $\pi^{\pm}$, 
$K^{\pm}$, $p$ and $\bar p$ interactions with protons. 
The solid lines represent fits of the data to a power law dependence. } 
\label{fig:differences}
\end{quote}
 \end{figure}

\begin{figure}[!h]
 \centering
 {\centering\resizebox*{!}{8cm}{\includegraphics{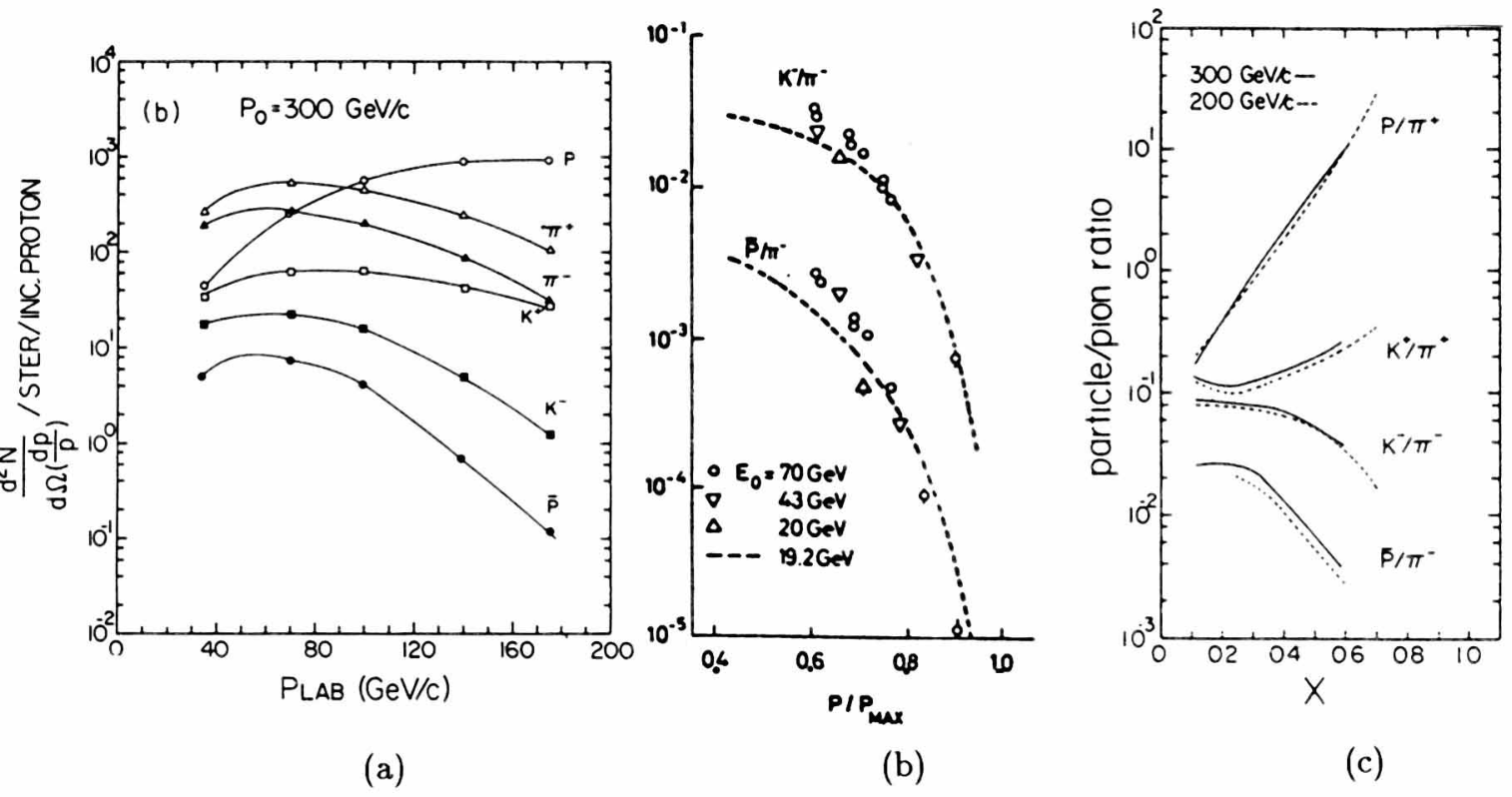}}\par}
\begin{quote}
 \caption{\small (a) Production cross sections of the six long-lived charged 
hadrons plotted vs lab momentum (incident proton beam $p_{lab}=300$ GeV/c); (b) (c )particle ratios 
vs $p / p_{max}$ at $p_{lab}$= 70 GeV at Serpukhov. } 
\label{fig:long}
\end{quote}
 \end{figure}

\section{Total cross sections at the CERN and Fermilab $\bar p p$ colliders}
The logical continuation of the total cross section measurements performed 
at the fixed target BNL, Serpukhov and Fermilab accelerators and at the CERN ISR $pp$ collider was to measure the total antiproton-proton cross section at the CERN \cite{bozzo, alner} 
and Fermilab \cite{amos, avila, abe} $\bar p p$ colliders, up to 1.8 TeV c.m. energy. 
A few members of the previous collaborations measured the antiproton-proton
 total and elastic cross sections at CERN and Fermilab. The CERN $\bar p p$ collider used a modified SPS, while the Fermilab collider used the superconducting ring. As already discussed for the CERN-ISR, at a $\bar pp$ collider, one needs a layout considerably different from the transmission measurements performed at fixed target accelerators. In order to measure elastic scattering at very small angles, precise detectors had to be positioned very far from the collision point ($\sim$100 m at Fermilab) inside containers (the so called ``Roman pots") placed very close (few mm) from the circulating beams. \par 
 
Since the circulating $p$ and $\bar p$ were inside the same ring, one could not use the Van der Meer method for measuring the luminosity: L is here known to a considerably smaller precision. Thus the luminosity independent method is more precise. \par  

Both the CERN and the Fermilab collider results established that the antiproton-proton 
total cross sections increase with increasing c.m. energies, Fig. 4b. The same experiments allowed to measure the high energy parameters: the total cross section $\sigma_{tot}$, the elastic cross section $\sigma_{el}$, the ratio $\sigma_{el} / \sigma_{tot}$, the parameter 
$\rho_{t=0}$, the slope B of the elastic nuclear differential cross section \cite{goulianos}.

\section{$pp$ total cross sections from cosmic rays} 
The cosmic ray $pp$ total cross sections shown in Fig. 4b were obtained in a rather indirect way. Cosmic ray exstensive air showers (EAS) measure the electromagnetic showers originated by p-air interactions yielding $\pi^0$ production followed by 
$\pi^0 \rightarrow 2 \gamma$ decay. EASs measure the attenuation $\Lambda$ of the rate of showers at different depths in the atmosphere. From this, the p-inelastic cross section may be obtained. A series of Monte Carlo simulations allow to correlate primary cosmic ray energy spectra, to interactions in air, electromagnetic showers, $\Lambda$ and $\sigma^{p-air}_{inel}$.
Other MCs correlate $\sigma^{p-air}_{inel}$ with $\sigma^{pp}_{tot}$, usually assuming the Glauber theory of p-air interactions \cite{glauber, pryke}. \par

The results have large statistical and systematic uncertainties. But nevertheless the data are in good agreement with increasing $pp$ total cross sections \cite{yodh}.\par

Analyses of the global $\bar p p$ and $pp$ data (measured by the Serpukhov, E104, ISR, UA4, UA5, CDF, E710, E811, cosmic ray experiments) using Regge pole formulae yield the following value for the $pp$ total cross section at the LHC: $\sigma^{pp}_{tot} (\sqrt s = 14  ~ TeV) \sim 108$ mb \cite{pryke} \cite{bloch}. \par

George Zatsepin was a pioneer in this field: he was a theoretician, a phenomenologist and performed experiments. He discussed the famous GZK cut off in the cosmic ray primary flux at $\sim 3 \cdot 10^{19}$ eV due to the interactions of the highest energy cosmic rays with the cosmic microwaves background radiation at 2.7 K. He determined several analytic formulae and he was the first to establish the chain: \\ 
CR spectrum $\rightarrow$ CR interactions with the atmosphere $\rightarrow$ production of $\pi^{\pm}$, $\pi^0 , K$ 
$\rightarrow$ $\pi^{\pm}$, $\pi^0 , K$ decays $\rightarrow$ EASs $\rightarrow$ $\Lambda$ $\rightarrow$ $\sigma_{tot}$ ...  \cite{zatsepin}. 

\section{Conclusions. Perspectives} 
In 1971 the experiment at the Serpukhov accelerator revealed that the $K^+$p total cross section increased with energy. In 1972 were published the first CR indications for rising $p p$ total cross sections. In 1973 the increase of the $pp$ total cross section was observed at the CERN ISR. Later in 1974-1978 the rising of $\pi^{\pm}$p, $K^-$p cross sections was observed at the Fermilab fix target accelerator and finally the rising $\bar p p$ was measured at the CERN and Fermilab $\bar p p$ colliders. We now know that at very high energies all total hadron-hadron cross sections increase with energy; this was confirmed, even if with lower precision, by the highest  energy cosmic ray 
data. (Also the $\gamma p$ total cross sections increase with energy). \par

From a theoretical point of view we still cannot obtain from the QCD lagrangian the answer to the question of why all the hadronic total cross sections grow with energy. In many QCD inspired models the rise may be connected 
with the increase of the number of minijets and thus to semi-hard 
gluon interactions. At the same time it is more or less clear that the rise of the total cross   sections is just the shadow of particle production: through the optical theorem the total cross section is related to the imaginary part of the elastic scattering amplitude in the forward direction. \par 

The high energy elastic and total cross section data vs energy have been 
usually analized in terms of Regge Poles, and thus in terms of Pomeron exchange. 
Even if the Pomeron was introduced long time ago 
we do not have a consensus on its exact definition and on its detailed 
substructure. Some authors view it as a ``gluon ladder''. From these fits were obtained predictions for the $p p$ total cross section at the LHC.\par 

Future experiments on hadron-hadron total cross sections will rely on the RHIC collider at BNL and mainly on the LHC proton-proton collider at CERN. \par 

Large area cosmic ray experiments may be able to improve the data in the ultra high energy region and solve some of the open problems, in particular the GZK cut off.\\

\vspace{0.5cm}

{\bf Acknowledgements.} I am greatful to all the collaborators in the various 
total cross section measurements. I thank Ms. Anastasia Casoni for typing the manuscript, Roberto Giacomelli and Miriam Giorgini for technical support.


\begin{thebibliography}{99}

\bibitem{diddens}  A. N. Diddens et al., Phys. Rev. 132 (1963) 2721. A. Citron et al.,  Phys. Rev. Lett. 13 (1964) 205. W. F. Baker et al., Phys. Rev. 129 (1963) 2285. 
W. Galbraith et al., Phys. Rev. 138 (1965) B913.
\vspace{-0.2cm}
\bibitem{cool} R. L. Cool et al.,  Phys. Rev. Lett. 16 (1966) 1228; Phys. Rev. Lett. 17 (1966) 102;  Phys. Rev. D1 (1970) 1887.
\vspace{-0.2cm}
\bibitem{abrams} R. J. Abrams et al., Phys. Rev. Lett. 18 (1967) 1209; 
Phys. Rev. Lett. 19 (1967) 259; Phys. Rev. Lett. 19 (1967) 678; Phys. Lett. 30B (1969) 564; Phys. Rev. D1 (1970) 1917; Phys. Rev. D1 (1970) 2477.
\vspace{-0.2cm}
\bibitem{gg} G. Giacomelli et al.,  Nucl. Phys. B37 (1972) 577; Nucl. Phys. B71 (1974) 138; Nucl. Phys. B20 (1970) 301.
\vspace{-0.2cm}
\bibitem{oh} Y. Oh et al., Phys. Rept. 423 (2006) 49.
\vspace{-0.2cm}
\bibitem{allaby} J. V. Allaby et al., Phys. Lett. 30B (1969) 500; Yad. Fiz. 12 (1970) 538.
\vspace{-0.2cm}
\bibitem{denisov} S. P. Denisov et al., Phys. Lett. 36B (1971) 415; Yad. Fiz. 14 (1971) 998. 
\vspace{-0.2cm}
\bibitem{cheng} H. Cheng et al., Phys. Rev. Lett. 24 (1970) 1456. J. Finkelstein et al., Phys. Lett. B34 (1971) 631.
\vspace{-0.2cm}
\bibitem{okun} L. B. Okun and I. Ya Pomeranchuck, Soviet Phys. JETP 3 (1956) 307.\\
I. Ya Pomeranchuck, Soviet Phys. JETP 7 (1958) 499.
\vspace{-0.2cm}
\bibitem{baker} Yu. B. Bushnin et al., Phys. Lett. 29B (1969) 48; Yad. Fiz. 10 (1969) 585. W. F. Baker et al., Nucl. Phys. B51 (1974) 303; Fermilab-78/79-EXP (1978).
\vspace{-0.2cm}
\bibitem{gg2}  G. Giacomelli, Progress in Nuclear Physics 12 (1970) 77; Nuovo Cimento Suppl. 6 (1968) 517; \\
Riv. Nuovo Cim. 2 (1970) 297.
\vspace{-0.2cm}
\bibitem{binon} F. Binon et al., Phys. Lett. 31B (1970) 230; Yad. Fiz. 12 (1970) 806. A. S. Carroll et al., Phys. Lett. B80 (1979) 319. R. J. Abrams et al., Phys. Rev. D4 (1971) 3235.
\vspace{-0.2cm}
\bibitem{yodh} G. B. Yodh et al, Phys. Rev. Lett. 28 (1972) 1005.\\
D. Cline et al.,  Phys. Rev. Lett. 31 (1973) 491.
\vspace{-0.2cm}
\bibitem{amaldi} U. Amaldi et al., Phys. Lett. B44 (1973) 112; Phys. Lett. B62 (1976) 460; Nucl. Phys. B145 (1978) 367. \\
S. R. Amendolia et al., Phys. Lett. B44 (1973) 119.
\vspace{-0.2cm}
\bibitem{gg3} G. Giacomelli, Phys. Rept. 23C (1976) 123.
\vspace{-0.2cm}
\bibitem{gg-jacob} G. Giacomelli and M. Jacob, Phys. Rept. 55 (1979) 1. \\
G. Giacomelli et al., Nuovo Cimento 24 C (2001) 575, hep-ex/0010070.
\vspace{-0.2cm}
\bibitem{carroll2} A. S. Carroll et al., Phys. Rev. Lett. 33 (1974) 928; Phys. Rev. Lett. 33 (1974) 932; Phys. Lett. B61 (1976) 303; Phys. Lett. B80 (1979) 423.
\vspace{-0.2cm}
\bibitem{antinucci} M. Antinucci et al., Lett. Nuovo Cimento 6 (1973) 121. \\
E. Albini et al., Nuovo Cim. A32 (1976) 101. \\
A. Breakstone et al., Phys. Rev. D30 (1984) 528. 
\vspace{-0.2cm}
\bibitem{bozzo} M. Bozzo et al., Phys. Lett. B147 (1984) 392. C. Angier et al., Phys. Lett. B344 (1995) 451.
\vspace{-0.2cm}
\bibitem{alner} G. J. Alner et al., Phys. Rept. 154 (1987) 247; Z. Phys. C32 (1986) 153.
\vspace{-0.2cm}
\bibitem{amos} N. A. Amos et al., Phys. Rev. Lett. 63 (1989) 2784; Phys. Lett B247 (1990) 127; Phys. Lett. B243 (1990) 158; Phys. Rev. Lett. 68 (1992) 2433; Nuovo Cimento 106A (1993) 123; Phys. Lett. B301 (1993) 313.
\vspace{-0.2cm}
\bibitem{avila} C. Avila et al., Phys. Lett. B445 (1999) 419; Phys. Lett. B537 (2002) 41.
\vspace{-0.2cm}
\bibitem{abe} F. Abe et al., Phys. Rev. D50 (1994) 5550.
\vspace{-0.2cm}
\bibitem{goulianos} K. Goulianos, hep-ex/07071055.
\vspace{-0.2cm}
\bibitem{glauber} R. J. Glauber, Lectures in Theoretical Physics, Interscience, NY (1956); R. J. Glauber et al., Nucl. Phys. B21 (1970) 135.
\vspace{-0.2cm}
\bibitem{pryke} C. L. Pryke, astro-ph/0003442 (2000). \\
R. S. Fletcher et al., Phys. Rev. D50 (1994) 5710. \\
A. M. Hillas, Nucl. Phys. B (Proc. Suppl.) 528 (1997) 29\\
J. Perez-Peraza et al., hep-ph/0408086.
\vspace{-0.2cm}
\bibitem{bloch} M. M. Block and R. N. Cahn, Rev. Mod. Phys. 57 (1985) 563; Phys. Lett. B188 (1987) 143.
\vspace{-0.2cm}
\bibitem{zatsepin} G. T. Zatsepin, Doklady Acad. Nauk. SSSR tom 99 N3 (1954) 369; Doklady Acad. Nauk. SSSR 67 (1949) 993; JETP 19 (1949).






\end{thebibliography}
\end{document}